# One step growth of PVP spheres embedded with nearly monodispersive CdS nanocrystals using chemical bath deposition

**Alka A Ingale[1], Rahul Aggarwal[1], Ekta Rani[1], Komal Bapna[1+], Pragya Tiwari[2] and Arvind K Srivastava[2]**

[1]Laser Physics Applications Section, Raja Ramanna Centre for Advanced Technology, Indore 452013; India. email id: alka.ingale@gmail.com

[2]Indus Sychrotron Utilisation division, Raja Ramanna Centre for Advanced Technology, Indore 452013; India.

**Abstract.** We have used simple chemical bath deposition technique to grow nearly monodispersive CdS nanocrystals in PVP matrix. Systematic study of variation of growth parameters has revealed that optimized growth of CdS nanocrystals in PVP matrix depends on relative concentration of Cd acetate/Thiourea to polyvinyl pyrrolidone in the bath. It is also observed that higher concentration (1M) of Cd acetate/Thiourea gives rise to smaller NCs compared to lower concentration (0.5M), however density of particles is large in thin film grown using 1M concentration. Scanning electron microscopic studies show that it's a nanoparticulate film of spheres of size ~ 100-200nm. Further, absorption, energy dispersive spectroscopy and transmission electron microscopic investigations reveal that nearly monodispersive CdS nanocrystals are embedded in 100-200 nm PVP spheres for the range 0.5 M, 1M Cd acetate/Thiourea concentration (figure 1). The effect of varying PVP, Cd acetate/Thiourea concentration, sequence and addition of ingredients and heating/ cooling cycles have been studied and results are corroborated with existing theory.

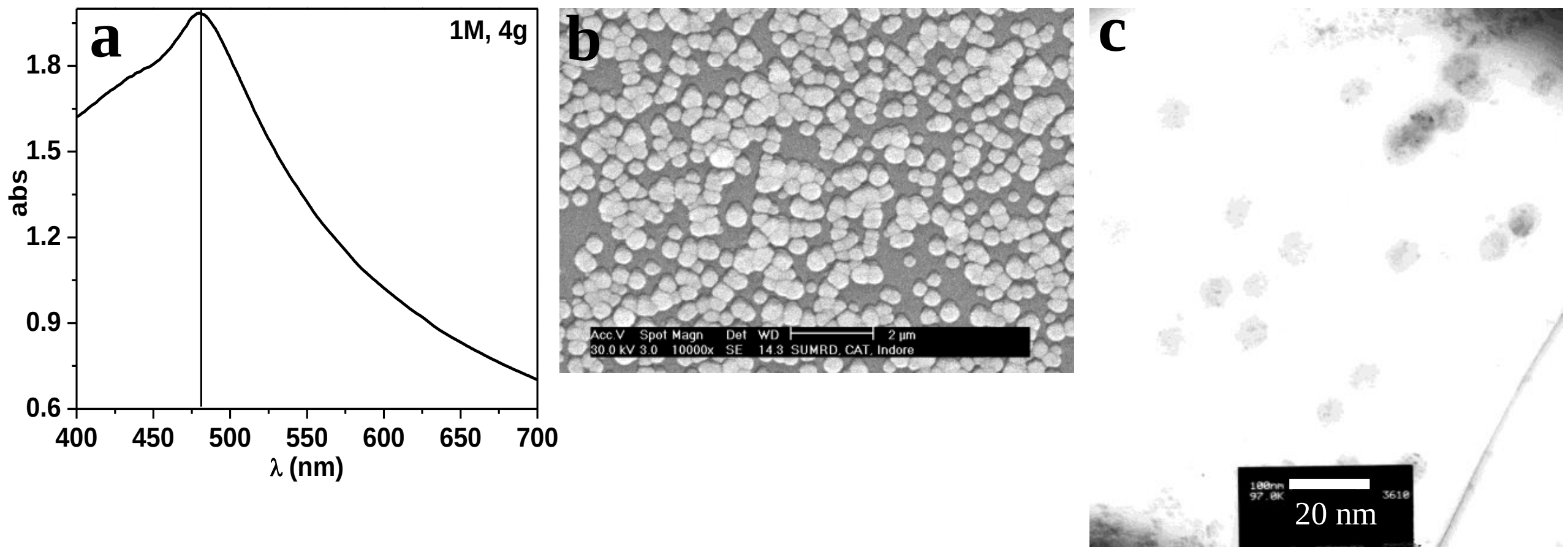

*Figure 1. (a) Absorption spectrum for CdS-PVP nanocomposite with 1M (Cd-A/Th) concentration and 4g PVP content, (b) SEM image of the nanocomposite film showing PVP spheres and (c) TEM image at the edge of one such PVP sphere showing embedded CdS nanoparticles.*

## I. Introduction

Semiconductor nanocrystals (NCs) show unique size dependent optical properties due to quantum confinement effect. Nanocomposite of inorganic nanostructures and organic polymer can effectively combine properties of both the components to create a material with new unique properties. This is highly desirable in order to exploit the full potential of the technological applications of the nanomaterials. For its use in microelectronics, sensors, optoelectronics etc., it is important to have thin films of composite materials. This has lead to extensive research in last decade in finding out suitable technique to create nanocomposites of semiconductor NCs embedded in polymer. Polymers provide processibility and mechanical strength keeping intact the optoelectronic properties of semiconductor NCs. In addition it provides steric hindrance and chemical passivation leading to NCs generation useful for optoelectronic applications. Finding new routes for driving organic polymers and inorganic NCs to self-assemble into nanocomposites is one of the challenges. Several groups are working towards this.[1-11] Another critical challenge is in predicting the morphology of these self-assembled hybrid materials. Balazs group is working towards this.[12] Experimental and theoretical work together can allow optimization of several alternate routes for incorporating semiconductor NCs in polymer matrix.

Cadmium sulfide (CdS) is an important semiconducting material, having unique optoelectronic properties as it is possible to engineer the band gap over a wide spectral range (visible to UV). It has been reported earlier that polyvinyl pyrrolidone (PVP), a water soluble polymer can effectively passivate the surface of NCs as well as act as matrix, thus restricting the agglomeration of NCs. It has been reported that quantum sized CdS NCs has promising third order non linear optical (NLO) properties as compared to bulk CdS and are useful for optical switches.[13] Embedding CdS NCs in a suitable polymer like polydiacetylene[14] and polystyrene[15], one can efficiently utilize the NLO property of NCs. Although, CdS-polymer nanocomposite has been relatively well investigated for various properties, systematic understanding of growth is not extensively explored this.[1-11] In this work we have done systematic variation of growth parameters to obtain interesting results which also throws light on growth mechanism of the nanocomposite.

In this paper we present a novel and simple method to create a thin film of CdS NCs embedded in PVP matrix on glass. There are several chemical methods for fabrication of semiconductor NCs either in powder form or thin film form.[16-21] One simple technique is chemical bath deposition (CBD). For the growth of NCs by CBD, two techniques can be employed (a) simple CBD method which involves direct mixing of metal and chalcogenide ions through some complexing agent[22] and (b) Modified CBD method which involves layer by layer adsorption of two opposite polarity species.[23] Here we report the growth details using first method, which we have also used earlier for growth of bulk CdS thin films of high optical and crystalline quality.[24] In this work, it has been studied that optimization of growth parameters needed for this growth are different from the best growth conditions obtained for bulk CdS growth.

The section 2 describes the basic experimental procedure involved in the growth of CdS NCs in PVP matrix. In the section 3, results obtained from scanning electron microscopy (SEM), absorption, Raman and transmission electron microscopy (TEM) measurements performed on grown samples are discussed. The observations of growth with variable bath parameters are explained in view of relevant theoretical studies reported in the literature in the section 4. The understanding of growth using the above is summarized in the section 5.

## II. Experimental

### A. Growth

The chemical bath used for preparation of nanocrystalline CdS thin films in PVP matrix is prepared in the following manner: Water bath is prepared by filling tap water in a large beaker and a smaller beaker is used for the chemical bath. The heating and stirring arrangement is made in the large and small beaker respectively. Chemical bath contained aqueous solution of PVP, Cd acetate (Cd-A: $Cd^{+2}$ source) and Thiourea (ThU: $S^{-2}$ source). Ammonia solution and Triethyl amine (TEA) are used as complexing agents to control number of $Cd^{+2}/S^{-2}$ and pH of the solutions. Before deposition, beakers and glass substrates are first cleaned. All the required beakers and glass slides are dipped in aqua-regia and left for 1 hr. Then all the equipments are washed in tap water and ultrasonically cleaned first in ethanol for 3-4 minutes and then in de-ionized water for same duration. After that, in order to prepare PVP solution, the measured amount of PVP is added to deionized water (105 ml) of this (chemical bath) beaker. Required number (2/3) of slides is dipped in it and the temperature is allowed to reach to as per the requirement. A stirrer is put in this solution which can rotate about its axis. Then in 25 ml of deionized water is taken in two small beakers separately and required weight of thiourea and cadmium acetate are added to these two beakers. TEA and liquid ammonia of required amount is added. Now the prepared ingredients are added in the required sequence.

To study growth of CdS NCs in PVP matrix from various angles, we have investigated,

(1) The effect of different molarities of $Cd^{+2}$ $S^{-2}$: Cd-A/ThU concentration varied from 0.01 to 1M in discrete steps.

(2) The effect of different quantity of PVP: quantity of PVP is varied from 0.1gm to 10g in discrete steps for 1M Cd-A/ThU.

(3) The effect of sequence of addition of ingradients like PVP, cadmium acetate and thiourea: with Cd Concentration 0.01M and 4gm of PVP.

(4) The growth of PVP: CdS-PVP nanocomposite growth is monitored with varying sequence of addition of ingradients, heating and cooling process of the chemical bath and duration of growth.

At various stages of development, cleaned glass slides are kept vertically in the chemical bath to have deposition for various durations. In all cases Cd-A and ThU concentrations are kept same in the chemical bath.

**B. Characterization**

Each thin film deposited is characterized using absorption spectroscopy with glass slide as reference using UV-3101PC spectrophotometer. Some of the chosen films are studied using scanning electron microscopy (SEM), energy dispersive spectroscopy (EDS), Raman and photoluminescence (PL) spectroscopy to get the further insight into the grown nanocomposites. For the same purpose transmission electron microscopy (TEM) is performed for one film. SEM images and EDS measurements are performed using model XL30CP (30KV) of Philips, Holland. The Raman spectra is recorded using double monochromator U1000 (Jobin-Yvon) and R649 (Hamamatsu) photomultiplier tube at room temperature in backscattering geometry.

## III. Results

There have been used many different methods like sol-gel, solvothermal methods to incorporate semiconductor NCs in polymer matrix, wherein semiconductor NCs are generated by other methods like micro emulsion, colloidal solution growth etc. and then used above methods to incorporate these NCs in Polymer matrix. In our method we use chemical bath containing all the required ingradients, wherein polymer is used not only to restrict the growth of semiconductor to form a NC by capping but also to embed NCs in the polymer matrix. We decided to start the work from the growth conditions optimized for bulk CdS.[23] We started with 4 gm of PVP and (0.01M Cd-A/ThU) we could get slight blue shift of band edge leading to average diameter size ~100-200 Å. Hereafter, we realized that to understand the growth mechanism, it is advisable to change variable parameters in a selected range, check their result and continue going in a direction of optimization. CdS-PVP growth is optimized using parameters like, 1) Cd-A concentration, 2) PVP concentration, 3) sequence of additions of chemicals and 4) Enhancing growth of PVP by heating, stirring etc. Soon, we realized that 1) The growth of PVP and CdS together is a co-operative phenomena and hence optimization of Cd-A/Thiourea concentration depends on PVP concentration. 2) The temperature at which the best CdS (thin film) growth could be obtained, the PVP growth is not favored at this temperature. This leads to larger CdS NCs. The details of the results of variation of different parameters are discussed below.

**A. Effect of different molarities of Cd-Acetate/Thiourea**

By varying Cd-A/ThU concentration from 0.01 M to 1M in discreet steps, we observe very interesting results for two concentrations 0.5M and 1 M, which are studied further for various growth conditions. These are studied for two PVP quantities 2g and 4g and CdS NCs growth is monitored further with different growth conditions (i.e. heating and cooling cycles). Total 10 such samples are studied. It is found that irrespective of the growth conditions chosen or the amount of PVP taken (i.e. 2g/4g), 0.5M concentration of Cd-A leads to larger size NCs compared to that for 1M Cd-A. The absorption spectra of few films are shown in figure 1a. Arrows indicate position of the band gap related peak. The observation of the peak, we believe is due to nearly monodispersive size distribution of CdS NCs obtained using these growth conditions. To substantiate this claim, we have performed photoluminescence (PL) of these samples. Figure 1b shows, absorption and PL spectra of sample S3. It is evident from the PL spectrum peaking at 468nm for absorption peak at 462nm that the absorption peak observed is indeed due to band gap. Calculated size of NCs from the blue shift of the band gap due to confinement[25] for S3 comes out ~65 Å. The absorption data indicates that for larger molarity of Cd-A/ThU smaller NCs are generated irrespective of variation in other parameters within the range specified above. This is not obvious, as one would expect faster growth of CdS with higher concentration of Cd-A/ThU (same amount of PVP) and hence larger CdS NCs. However, in accordance with higher concentration, it is found that absorbance has higher value for higher Cd-A/ThU concentrations suggesting larger density of NCs in this case.

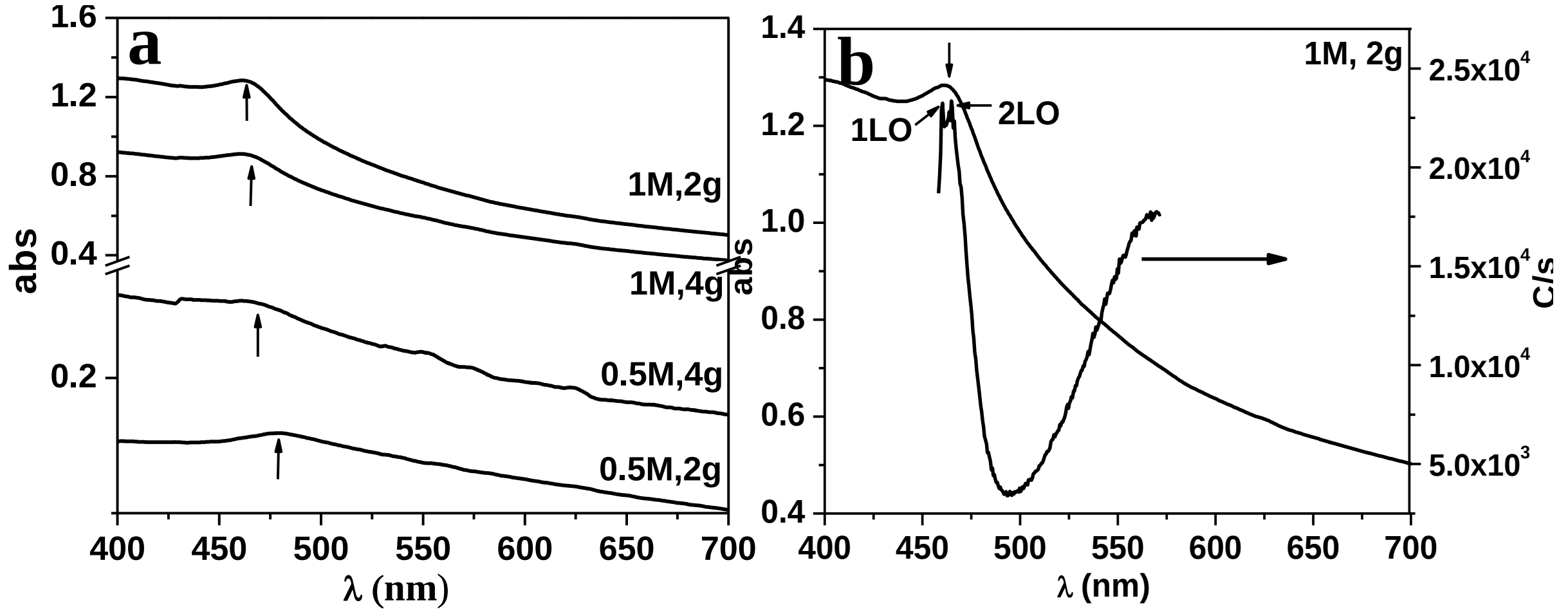

Figure 1. (a) Absorption spectra of samples with different concentration of Cd-A/ThU and PVP as shown on the top of plot and (b) Absorption and PL spectrum of sample with Cd-A/T and PVP concentration as 1M/2g.

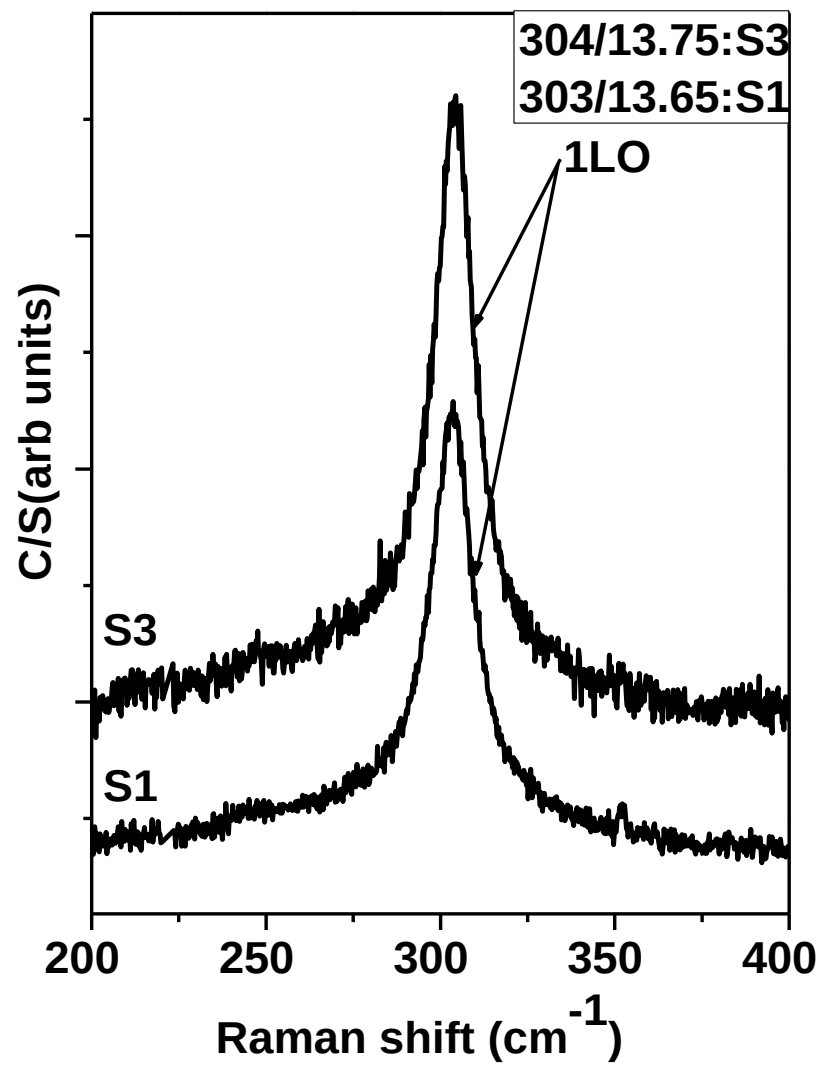

Figure 2. Raman spectra of samples S1 and S3.

Raman measurements are performed on few samples to confirm formation of CdS. Representative Raman spectra is given in figure 2. Detailed analysis of Raman measurements will be published elsewhere.[26]

### B. Effect of PVP concentration

PVP quantity viz 0.05, 0.1,1, 2, 4 and 10 g is used in the 9 samples studied. It is found that reduction in concentration of PVP from 10 gm to 0.05 gm, keeping the Cd-A/ThU concentration same (0.5M), allowed the growth of larger CdS NCs (figure 3). Narrow absorption peaks are observed at 480nm (0.05gm PVP, S5), 480nm (0.1gm PVP, S6) and 478nm (2gm PVP, S8 (same as S1)) indicating the near monodispersity of CdS NCs sizes. However, absorption spectrum of S10 with 10gm PVP shows smaller (452 nm) CdS NCs with larger size distribution, as absorption edge is observed (figure 3). Since more the PVP, faster is the growth of PVP matrix which, through better capping, breaks the development of larger CdS particles, and hence leading to smaller NCs. Almost similar particles sizes are observed for 4gm, 2gm, 0.1gm, 0.05gm PVP and significant change in size is observed for 10gm of PVP. It is interesting to note that sample S5 with 0.05 gm PVP has larger absorbance peak value in comparison to S10 with 10gm PVP, consistent with larger Cd-A/ThU concentration w. r. t. PVP. For S10 with 10 gm PVP, low scattering (figure 3: absorbance ~700nm) is observed as compared to other samples (S8, S6, S5) with smaller PVP concentration, which shows the more homogenous growth of PVP matrix in samples with larger PVP concentration. Absorption spectra of two other samples S11 (1M Cd, 2g PVP) and S12 (1M Cd, 4g PVP) shown in figure 4a shows the similar absorption peaks at 496nm and 499nm respectively, indicating similar particle sizes. Both the samples show similar scattering but sample S11 with 2 gm PVP exhibited larger peak absorbance value as compared to sample S12 with 4g PVP which is consistent with our earlier observations.

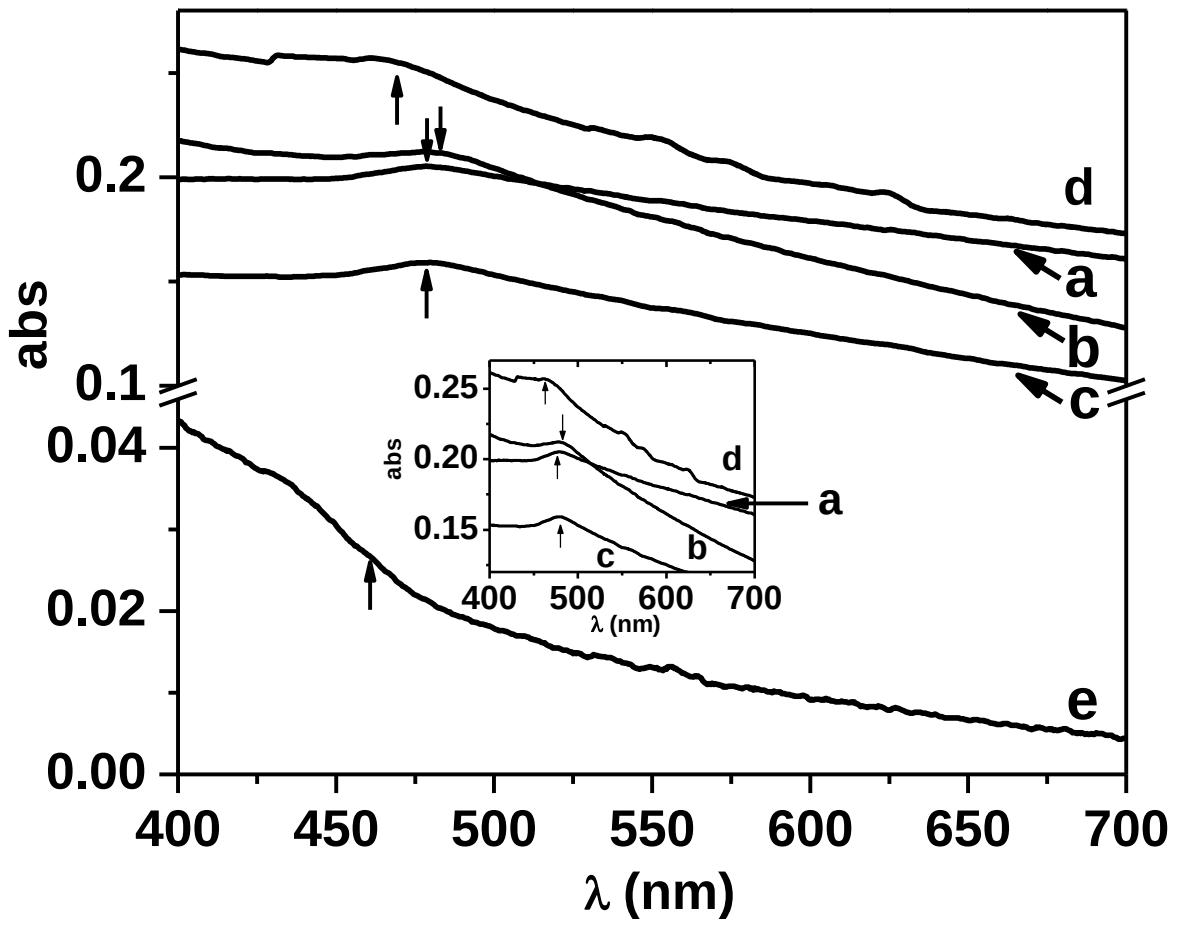

Figure 3. Absorption spectra of samples with Cd-A/ThU concentration as 0.5 M and different concentration of PVP as (a) 0.05g, (b) 0.1g, (c) 2g, (d) 4g and (e) 10g.

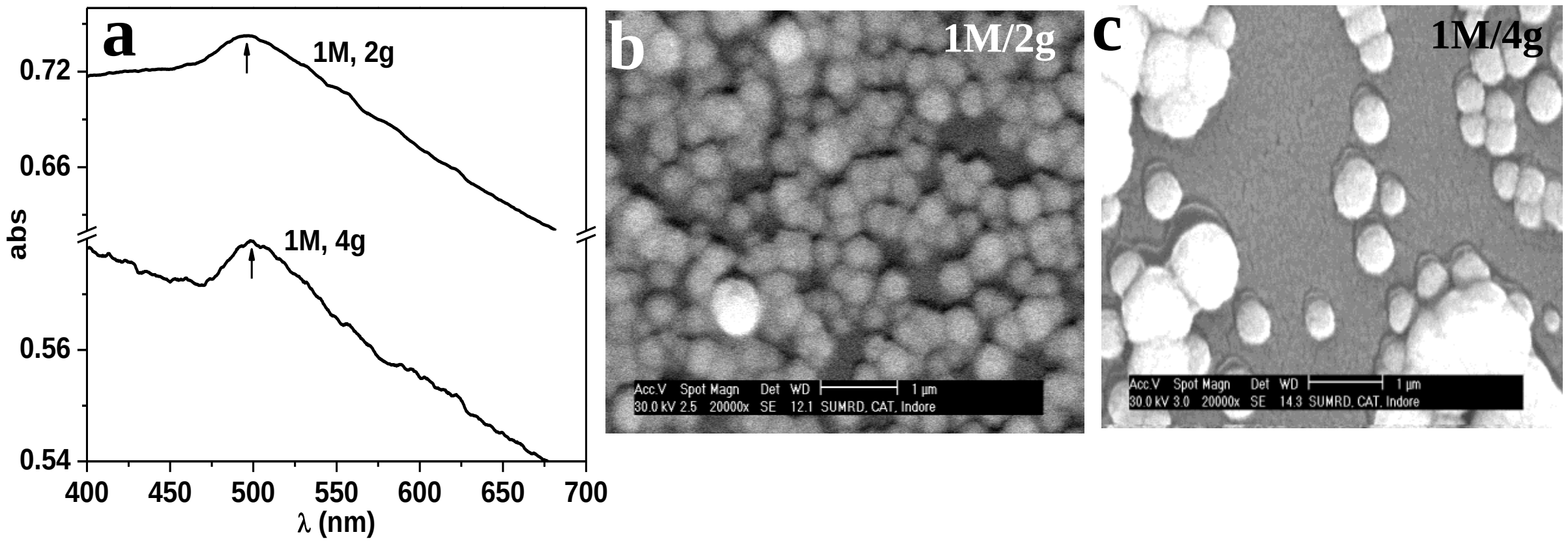

Figure 4. (a) Absorption spectra of samples with Cd-A/ThU concentration as 1 M and different concentration of PVP as noted in the plot, (b) & (c) SEM image of the same samples.

**C. Size of CdS NCs: consistency of absorption, SEM and TEM data**

SEM is performed to study morphology of the films grown. Representative data is shown in figure 4 for S11 and S12 along with absorption data. The difference between two samples is quantity of PVP. SEM pictures shows that S11 contains uniform distribution of white spheres, nearly monodispersive (~2000 Å) in size. However, S12 shows clustering of these spheres extensively, where larger PVP quantity is used. Similar observation is made for many sets of other samples too. This lead us to believe that these white spheres are polymer spheres and CdS NCs of appropriate sizes (~ 60-100 Å) would be in the black region. To confirm this we performed EDS measurements on these samples. We found appropriate stoichiometric ratio could be obtained for Cd and S only in the region of white spheres and no Cd/S could be seen elsewhere (black region). This is rather intriguing. The only possibility we could think of is that smaller size CdS NCs are embedded in these polymer spheres. To check this, we performed TEM of S13. S13 is chosen due to sharp absorption peak observed for the sample as shown in figure 5a. Figure 5b and c shows SEM and TEM of the same. For TEM, one white sphere is chosen and is looked into with high resolution. Since, e beam needed to cross lot of polymer material, the observation of CdS NCs shown are very difficult to achieve. *The TEM picture clearly shows near monodispersivity of the CdS NCs along with consistency of* *size obtained from absorption spectra (~8.5 nm).* This establishes the fact that we have achieved the growth of nearly monodispersive spherical CdS NCs embedded in polymer, which is sphere in themselves. It is interesting to understand why this morphology is formed and this will be discussed in section of theory corroboration, section 4.

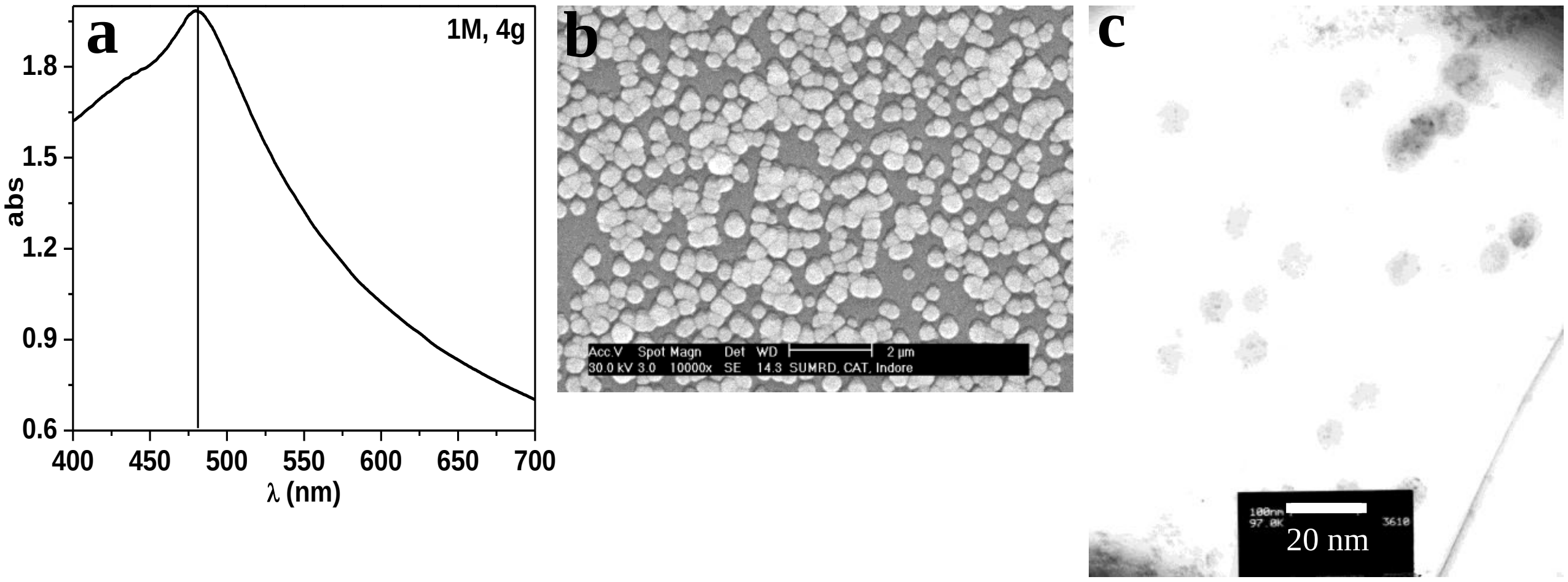

Figure 5. (a) Absorption spectra of sample with Cd-A/ThU and PVP concentration as 1M/4g, (b) SEM image and (c) TEM image of same sample.

**D. To understand the suitable conditions for the growth of PVP**

**a. Effect of deposition time.**

We have studied 10 samples with same PVP and Cd-A/ThU molarities by varying the deposition time. Absorption spectra of one such set are shown in figure 6a. S2, the sample taken out from same chemical bath after being subjected to heating cycle and additional deposition time is numbered S13. The similar procedure is followed for S11 also. Figure 6a shows that absorption peak red shifts for S2, S13, S11 indicating increase in size from 6.5nm to 12.5nm. It is important to note that size of NCs does not increase for heating cycle below 70 $^{0}$C. This suggests that polymer capping opens up to allow further growth of CdS before recapping by polymer takes place at RT. Polymer conformation is also indicated from the SEM pictures of S11, wherein longer deposition shows smaller polymer spheres compared to that of S13. It is also observed that longer deposition times lead to larger amount of scattering (absorbance ~500-700 nm) from polymer matrix due to inhomogeneous growth.

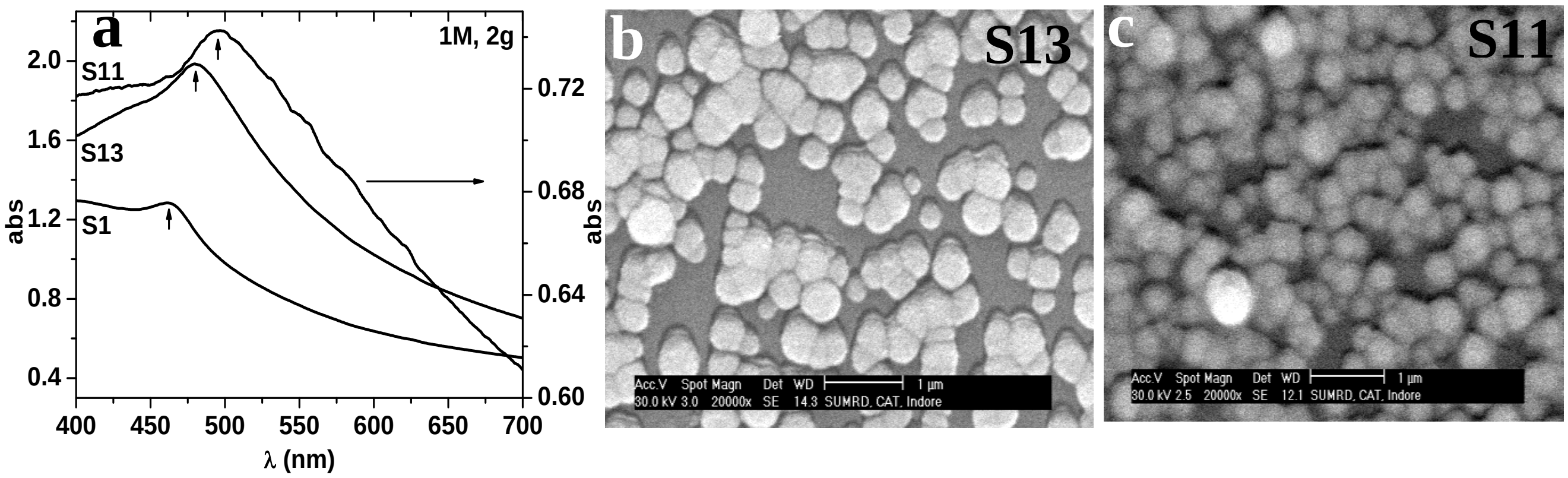

Figure 6. (a) Absorption spectra of samples with Cd-A/ThU and PVP concentration as 1M/2g and different growth conditions as noted in the text and (b&c) Comparison of SEM images of samples S13 and S11.

**b. Effect of time sequence of addition of ingredients.**

Growth of 10 samples are studied by changing sequence of addition and allowing different bath conditions for allowing prepolymerization as detailed in the following. PVP is added to 100 ml deionized water at 70 °C and then allowed to cool at room temperature for 1 hour. Time of addition of Cd-A and ThU is varied. For sample S14 and S15, Cd-A and ThU are added simultaneously at 70 °C but sample S15 is exposed to room temperature for 3 more hours.

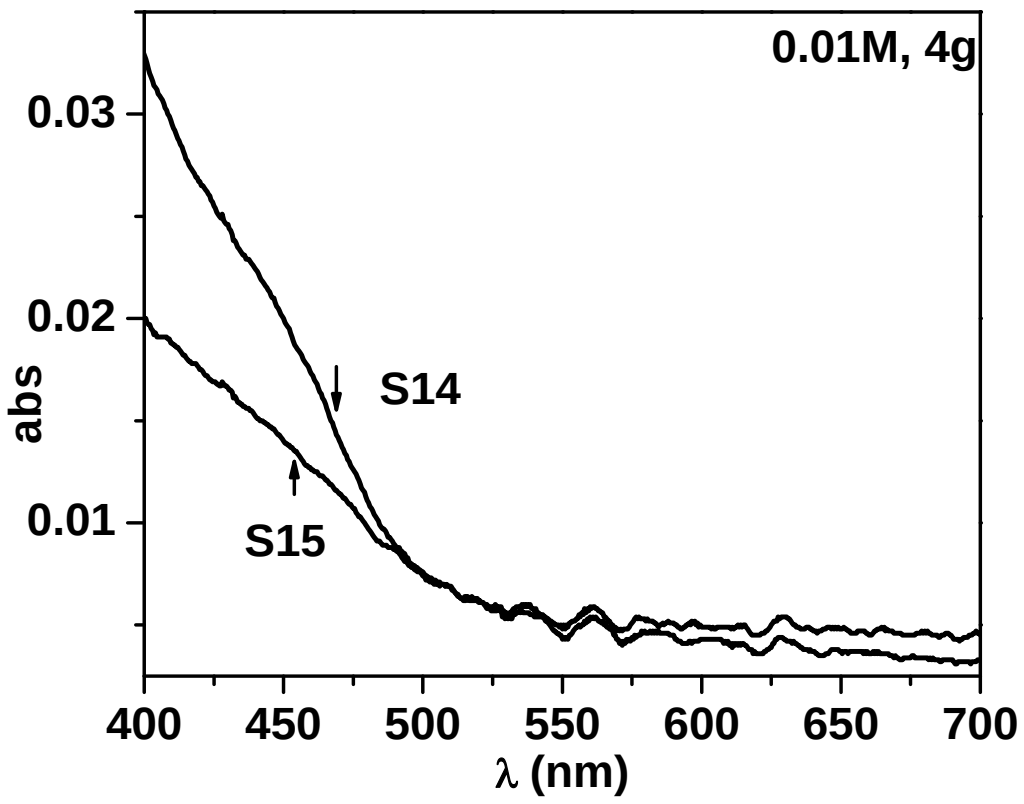

Figure 7. Absorption spectra of samples with Cd-A/ThU and PVP concentration as 0.01M/4g and different time of growth as noted in the text.

Absorption spectra shows blue shifted absorption edge at 453 nm (figure 7) for sample S15 in comparison to absorption edge at 468 nm (figure 7) for sample S14.This indicates that the polymerization is faster with respect to CdS growth at lower temperature consistent with earlier observations that temperatures higher than 70 $^{0}$C polymer undergoes conformational change.

In the next experiment (for sample S16 and S17) Cd-A is added at 80 °C (for half an hour) and left at room temperature for 16 hrs to get complete dissolution of Cd-A and then ThU is added and again heated the solution (note here thiourea is added after 16 hrs of RT) to 90 °C. Sample S17 is exposed to room temperature for 27 more hours. Absorption edge (figure 8) at around 485 nm is observed for sample S16 (no abs curve and Raman for S17). It is known to have good quality growth of CdS bulk film at 90 °C, it is found that CdS grows faster and therefore larger NCs are grown at 90 °C than compared to 70 °C.

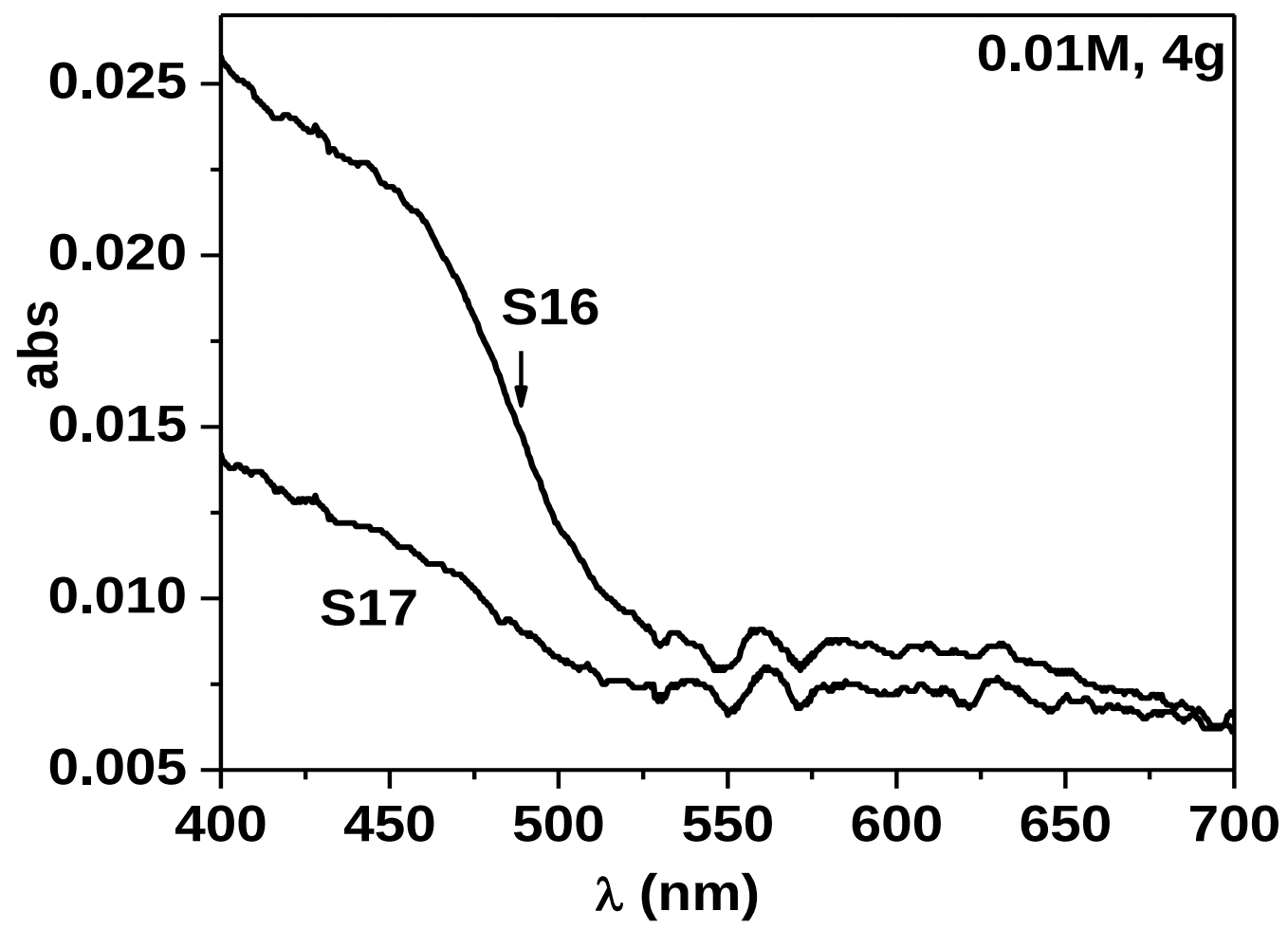

Figure 8. Absorption spectra of samples with Cd-A/ThU and PVP concentration as 0.01M/4g and different time of growth as noted in the text.

In the third experiment (sample S18-22 (figure 9)) PVP is allowed to polymerize for long (26 hours) at room temperature and then Cd-A and ThU are added at 70 °C. Sample S18-20 which is kept (after heating) at room temperature for 27 hours, 4 days and 5 days respectively gave continuous curve with no absorption edge. After 5 days the fresh slides are put in and bath is heated to 70 °C. Both the samples, S21 and S22 which are removed after heating at 70°C for 1hour and 2 hours respectively, registered the absorption edge at ~ 474 nm. This indicates that the long exposure to room temperature affected the formation of CdS NCs as polymerization has taken over the growth of CdS. Either very small particles are formed or CdS has not captured inside the PVP matrix.

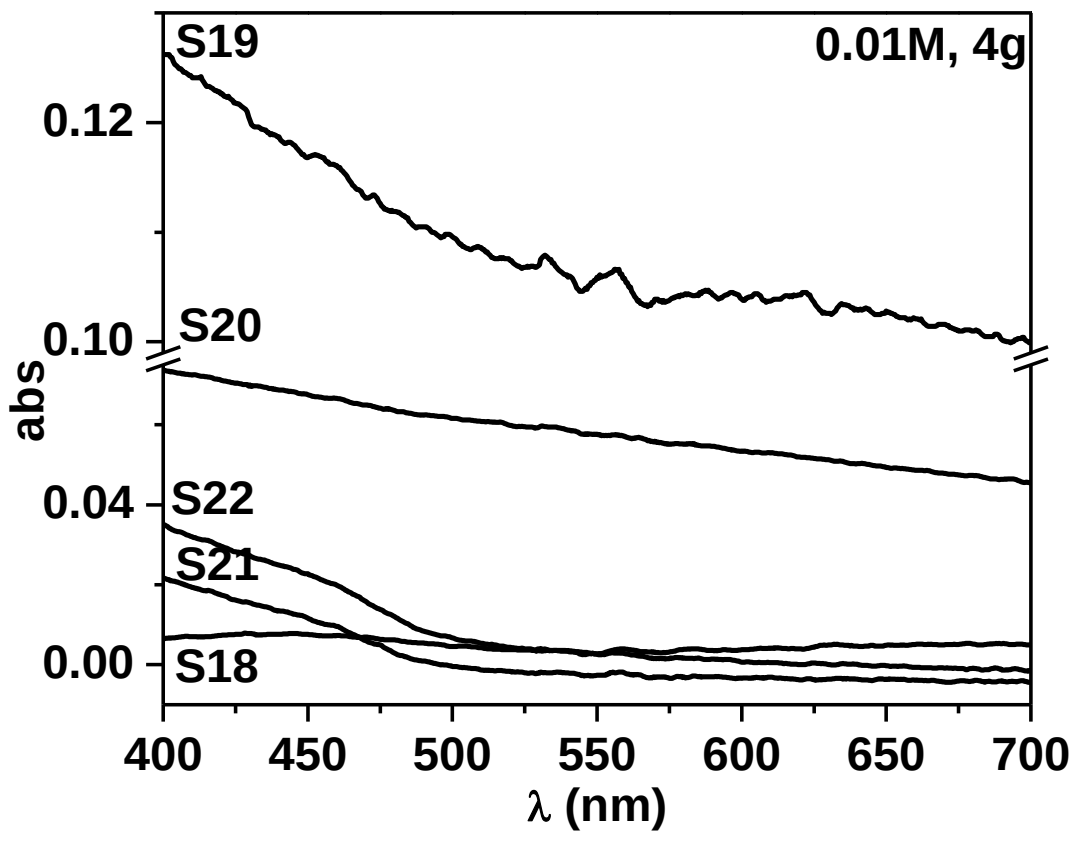

Figure 9: Absorption spectra of samples with Cd-A/ThU and PVP concentration as 0.01M/4g and different time of growth as noted in the text.

4g PVP is dissolved in 105 ml of deionised water in a beaker and is kept in a water bath. It is allowed to heat to 90 °C with constant stirring, so that PVP could get dissolved completely. Aqueous solution of like cadmium acetate, thiourea, tri-ethyl amine (TEA) and liquid ammonia are added to this solution in the same sequence. Five different sets are prepared to understand the process of polymerization. In the first set PVP is heated to 90 °C and then cooled to room temperature (sample S23, S24, S25). Absorption spectra registered the band edge at around 517nm indicating the formation of very large (bulk) CdS particles. For the second (sample S26, S27, S28), third ( sample S29, S30, S31) and fourth set (sample S32,S33,S34) PVP is heated till 90 °C and maintained at the same temperature for 5,15 and 30 minutes respectively. For second and third sets, absorption edges are noted at ~482 nm and for ~492 nm respectively indicating the blue shift in size of CdS particles. No absorption edge is registered for fourth set which indicates that CdS NCs have not been formed. By allowing smaller time for polymerization, better capping is resulted in smaller NCs, whereas, in larger (30 min) time, polymerization does not allow the growth of CdS NCs at all. In between condition, we find surprisingly larger NCs formation, which is not obvious to understand. In the fifth set (sample S35 and S36) PVP is heated till 90 °C and solvents are added immediately after which gives the absorption edge at 486 nm.

**E. Theory corroboration**

There are quite a few observations in this set of experiments which can be termed as unusual, since they cannot be explained using simple and common reasoning. Therefore, we look for possible explanation in the existing literature related to growth modeling using different theories. These results and qualitative corroboration obtained from literature is given below.

1. *Observation:* One expects CdS NCs to be embedded in two dimensional polymer matrix in CBD grown films. However, SEM and TEM results show that CdS NCs are embedded in three dimensional PVP matrix with matrix taking morphology of a sphere.

A collapse of a polymer in a binary solvent leading to formation of globules has been explained in terms of effective mediated attraction between the monomers.[27] Diamant et al. have shown that in addition to interaction between particle/molecule and polymer, intra-polymer interactions have an important role to play in partial collapse of the polymers.[28] In PVP, the pyrrolidone part has two coordination groups, N and C = O. Recently in PVP–Mn doped CdS nanocomposite, it has been observed that oxygen in carboxylates group makes coordinate bond with the Cd ions.[29] This is the evidence of attractive PVP-CdS NC interaction. This understanding explains formation of polymer spheres embedded with CdS NCs observed here. Further, it may be noted that the concentration of polymer and molecules/NCs is also a determining factor for condition of partial collapse of polymer.[28]

*Observation:* Further, our absorption data indicates decrease in size of CdS NCs for higher concentrations of Cd/S (0.5 and 1M) while keeping same PVP concentration (2g/4g).

Cao et al. has reported density functional theory (DFT) for a coarse-grained model of block copolymer-nanoparticle mixtures and predicts that well-organized structures can be effectively controlled by adjusting the polymer chain length and polymer-particle interaction.[30] In our case we have kept polymer chain length fixed (K40). Thus the polymer-particle interaction is the main parameter controlling the structure/morphology in our case. This polymer-particle interaction can be changed with change in concentration of PVP & or Cd/S ions, complexing agents, temperature and sequence of addition of solvents. These variations in parameters can give rise to different size CdS NCs and different morphology of CdS-PVP nanocomposites as noted earlier. To predict the equilibrium morphology of polymeric mixtures, the microphase separation of the copolymers into composites, Russell and co-workers has developed mean field theory for mixtures of soft, flexible chains and hard spheres. The model integrates a self-consistent field theory (SCFT) for polymers and a density functional theory (DFT) for particles.[31] Simulation, observes that in such a complex mixtures, it is not simply the ordering of the polymers that effect the spatial organization of the particles; the particles in fact affect the self-organization of the copolymers.[31] This co-operative growth mechanism can lead to lower size Cd/S NCs in higher concentration of Cd/S ion source which depends on relative PVP concentration, while conserving the total amount of CdS in the film as observed in our case.

2. *Observation*: While changing the ratio of CdS/PVP, we observed that there is optimum concentration of PVP for fixed concentration of Cd/S ions to have narrow size distribution of NCs.

Balazs has predicted that in nanocomposite, only NCs having sizes in a particular range will form. Polymer chains must stretch around these NCs, which require energy.[12] This loss increases with the increase in particle radius. Therefore larger NCs will not be included in the bulk of the polymers as it is not favored thermodynamically. Also the process of formation and dissolution of NCs is an equilibrium process and thus smaller particles may be dissolved. This will lead to monodispersity in size of NCs within polymers as is observed in our case.

In view of the above discussion it is interesting to note the following:

i) The CdS NCs size increases with increase in concentration of Cd/S ~ in mM (PVP: 2-4g) as observed earlier[4 & 9] and by us. This suggests that relation of size of CdS NCs and Cd-A/ThU concentration is itself a function of relative concentration of Cd/S and PVP confirming further that CdS-PVP nanocomposite growth is a co-operative growth.

ii) Heating / vigorous stirring actually leads to size distribution although it may be homogenous as reported.[7,10] This suggests that if additional KE is provided via heating or stirring, it allows larger size NCs to be included in polymer matrix thus leading to larger size distribution.

iii) 0.5-1 M Cd/S and 2-4g PVP gives rise to most interesting set of nearly monodispersive CdS NCs. Indicating suitable co-operative growth and polymer collapse conditions for these nanocomposites.

iv) *Elashmawi et al. 2013 have also reported decrease* in crystallite size as with increase in Cd/S concentration with Cd/S concentration in ~ 0.6, 1.2, 2.4 M (similar to our case). However, their observed size of CdS NC and calculated size from absorption spectra shows inconsistency, probably as the spheres observed in SEM has been identified as CdS NCs instead of polymer embedded with CdS NCs as is shown in this paper.[11]

## F. Conclusion

One step growth of CdS-PVP nanocomposites is performed using simple chemical bath deposition on glass slides. To get insight into growth mechanism the parameters of bath like concentrations of Cd-A(Cd source), thiourea(S source) and PVP are varied from 0.01 M to 1M and 0.01 to 10g respectively. The heating and cooling cycles also are varied to see how well PVP can restrict the growth of CdS under these conditions.

One of the important observations is that the optimum growth conditions for CdS thin film is not the same for CdS-PVP nanocomposite. Instead it is observed that optimum growth condition for the nanocomposite depends on the ratio of Cd-A(Cd source)/thiourea(S source) concentration to PVP content indicating it is a co-operative growth. The 0.01 M concentrations produce smaller size but large size distribution and the most interesting concentrations of Cd-A(Cd source)/thiourea(S source) are 0.5 and 1M with either 2 or 4 g PVP . This gives nearly monodispersive CdS NCs. The literature contains some results, wherein SEM observed images of polymer spheres has been credited as CdS NCs with incorrect size estimation from absorption. We have sorted the inconsistency by TEM imaging the edge of a sphere showing CdS nanoparticles of smaller dimensions matching the size that obtained from absorption data. Formation of PVP spheres, decrease in size and increase in density of CdS NCs with higher molar concentration of Cd-A (Cd source)/thiourea(S source) and nearly monodispersive CdS NCs embedded in PVP spheres are explained satisfactorily using existing theory and understanding.[27, 28 30 & 31] The quantitative understanding however will need theoretical modeling for the parameters in the range of interest. This is being explored at present. It is interesting to note that CdS NCs embedded in PVP matrix has shown promising and better NLO property as compared to CdS NCs embedded in other polymer matrices.[32] We plan to study NLO properties of these samples to understand effect of morphology of the nanocomposite on it's NLO properties.

**Acknowledgement:**

Dr. S. C. Mehendale, Dr. H. S. Rawat and Dr. S. K. Deb are greatfully acknowledged for support during this work. Ms Ekta Rani wishes to acknowledge Homi Bhabha National Institute, India for providing research fellowship during the course of this work.

**References**

[1]A. A. Patel, F. Wu, J. Z. Zhang, C. L. Torres-Martinez, R. K. Mehra, Y. Yang and S. H. Risbud, J. Phys. Chem. B **104**, 11598 (2000).

[2]R. He, X. F. Qia, J. Yin, H. Xi, L. Bian and Z. Zhu, Colloids and Surfaces A: Physicochem. Eng. Aspects **220**, 151 (2003).

[3]C. Lü, C. Guan, Y. Liu, Y. Cheng and B. Yang, Chem. Mater. **17**, 2448 (2005).

[4]M. Pattabi, B. Saraswathi and A. K. Manzoor, Materials Research Bulletin **42**, 828 (2007).

[5]Q. Xia, X. Chen, K. Zhao and J. Liu, Materials Chemistry and Physics **111**, 98 (2008).

[6]D. J. Asunskis, I. L. Bolotin and L. Hanley, J. Phys. Chem. C **112**, 9555 (2008).

[7]S. Peretz, B. Sava, M. Elisa and G. Stanciu, J. of Optoelectronics and Advanced Materials **11**, 2108 (2009).

[8]M. L. Singla, M. Shafeeq, M. Kumar, J. of Luminescence **129**, 434 (2009).

[9]D. S. Yoo, S. Y. Ha, I. G. Kim, M. S. Choo, K. Kim and E. S. Lee, New Physics: Sae Mulli **61**, 680 (2011).

[10]L. Saravanan, S. Diwakar, R. Mohankumar, A. Pandurangan and R. Jayavel, Nanomater. Nanotechnol. 1, 42 (2011).

[11]I. S. Elashmawi, A. M. Abdelghany and N. A. Hakeem, J Mater Sci: Mater Electron **24**, 2956 (2013).

[12]A. C. Balaz, T. Emrick, T. P. Rusell, Science **314,** 1107 (2006) & references therein.

[13]N. Venkatram, R. K. Sai Santosh and R. D. Narayana, J. Appl. Phys. **100,** 074309 (2006).

[14]R. E. Schwerzel, K. B. Spahr, J. P. Kurmer, V. E. WoodJerry and A. Jenkins, J. Phys. Chem. A **102,** 5622 (1998).

[15]H. Du, G. Q. Xu, W. S. Chin, L. Huang and W. Ji, Chem. Mater. **14**, 4473-4479 (2002).

[16]P. K. Nair, M. T. S. Nair, A. Fernandez and M. Ocampo, J. Phys. D: Appl. Phys. **22**, 829 (1989).

[17]C. Nascu, V. Vomir, I. Pop, V. Ionescu and R. Grecu, Mater. Sci. Eng. B **41**, 235 (1996).

[18]J. J. Valenzuela, R. Ramirez, A. Mendoza and M. Sotelo, Thin Solid Films **441**, 104 (2003).

[19]R. R. Hawaldar, G. G. Umarji, S. A. Ketkar, S. D. Sathaye, U. P. Malik and D. P. Amalnerkar Mater. Sci. Eng. B **132**, 170 (2006).

[20]R. S. Mane and C. D. Lokhande, Mater. Chem. Phys. **65**, 1 (2000).

[21]K. M. Gadave, S. A. Jodgudri and C. D. Lokhande, Thin Solid Films **245**, 7 (1994).

[22]R. S. Patil, C. D. Lokhande, R. S. Mane, T. P. Gujar and S. H. Han, J. of Non-Crystalline Solids **353**, 1645 (2007).

[23]V. Shukla, V. K. Dixit and A. A. Ingale Proceedings of DAE- Solid state Physics symposium **50**, 471 India (2005)

[24]A. A. Ingale, R. Aggarwal, K. Bapna, P. Tiwari and A. K. Srivastava ICONSAT-2010 p. 213 (2010).

[25]Y. Wang and N. Herron, Phys. Rev. B **42**, 11 (1990).

[26]Ekta Rani, Alka A. Ingale, (to be submitted)

[27]P. G. de Gennes, J. Phys. Lett. **37**, 59–61 (1976).

[28]H. Diamant and D. Andelman, Macromolecules **33**, 8050 (2000).

[29]P. Verma, G. S. Manoj, A. C. Pandey, Physica B **405**, 1253 (2010).

[30]D. Cao, J. Wu, The J. of Chem. Phys. **126**, 144912 (2007).

[31]R. B. Thompson, V. V. Ginzburg, W. M. Matsen and A. C. Balazs AC, Macromolecules **35**, 1060 (2002).

[32]C. Jing, X. Xu, X. Zhang, Z. Liu and J. Chu, J. Phys. D: Appl. Phys. **42,** 075402 (2009).